\documentclass[review]{elsarticle}

\usepackage{lineno,hyperref}
\modulolinenumbers[5]

\usepackage{times}
\usepackage{amsmath,amssymb}
\usepackage{epsfig,graphics}

\def\rd{{\rm d}}
\def\vx{{\bf x}}
\def\vy{{\bf y}}

\journal{Journal of \LaTeX\ Templates}

\begin{document}

\begin{frontmatter}

\title{Representations and
Divergences in the Space of Probability Measures and
Stochastic Thermodynamics}


\author[firstaddress,secondaddress]{Liu Hong}
\ead{zcamhl@tsinghua.edu.cn}

\author[firstaddress]{Hong Qian\corref{correspondingauthor}}
\cortext[correspondingauthor]{Corresponding author}
\ead{hqian@uw.edu}

\author[firstaddress]{Lowell F. Thompson}
\ead{lowell.f.thompson@gmail.com}

\address[firstaddress]{Department of Applied Mathematics, University of Washington, Seattle, WA 98195-3925, U.S.A.}
\address[secondaddress]{Zhou Pei-Yuan Center for Applied Mathematics, Tsinghua University, Beijing, 100084, P.R.C.}

\begin{abstract}
Radon-Nikodym (RN) derivative between two measures
arises naturally in the affine structure of the space of
probability measures with densities.  Entropy, free
energy, relative entropy, and entropy production as mathematical concepts associated with RN derivatives are introduced.
We identify a simple equation that connects two measures
with densities as a possible mathematical basis of
the entropy balance equation that is central in 
nonequilibrium thermodynamics.
Application of this formalism to Gibbsian canonical distribution
yields many results in classical thermomechanics.  An
affine structure based on the canonical represenation
and two divergences are introduced in the space of probability
measures.  It is shown that thermodynamic work,
as a conditional expectation, is indictive of the RN derivative
between two energy represenations being singular.
The entropy divergence and the heat divergence yield respectively
a Massieu-Planck potential based and a generalized
Carnot inequalities.
\end{abstract}


\end{frontmatter}


\section{Introduction}
	
	A subtle distinction exists between the prevalent
approach to stochastic processes in traditional 
applied mathematics and the
physicist's perspective on stochastic dynamics:  In
Kolmogorov's theory of stochastic processes, the dynamics
are described in terms of a trajectory $\{\vx(t): t\in[0,\infty)\}$.
Applied mathematicians treat each of these trajectories as a random event in a large probability space and then study probability
distributions over the space of all possible trajectories $\vx(t)$.
Physists, however, are more accustomed
to thinking of a ``probability distribution changing with time,''
$\rho(\vx,t)$.  In the case of continuous-time Markov
processes, $\rho(\vx, t)$ is described by
the solution to a Fokker-Planck equation or
a master equation, while for a Markov chain simply
by a stochastic matrix.  This latter perspective
can perhaps be more rigorously formulated in
the {\em space of probability measures}.  The dynamics are then represented as a {\em change of measure}.
The Radon-Nikodym (RN) derivative is a key
mathematical concept associated with changes of measure
\cite{kolmogorov-book}.  Interestingly, RN derivative between
two measures is also at the heart
of the concept of {\em fluctuating entropy}
\cite{qian-pre01b,ye-qian-18}.

	This ``probability distribution changing with time''
view is, of course, not foreign to mathematics.  Actually in
the 1950s, the stochastic diffusion process developed by Feller,
Nelson, and others was precisely a such theory \cite{feller,nelson,jqq,pavliotis-book}.
That approach, based on solutions to linear parabolic partial
differential equations, was formulated in a
linear function space.  We now know that a more geometrically
intrinsic representation for the space of probability
measures cannot be linear: There is simply no natural choice of origin.  Rather, an {\em affine space} is
more appropriate \cite{jeangallier, thompson}.

	Entropy and energy are key concepts in the classical
theory of thermodynamics, which is now well understood to
have a probabilistic basis.  In fact, one could argue that the
very notion of ``heat'' arises only when one treats the
motions of deterministic Newtonian point masses
as stochastic.  In the statistical treatment of thermodynamics,
Gibbs' canonical energy distribution is one of the key
results that characterize a thermodynamic equilibrim
\cite{pauli}.  As we shall see, it figures prominently in
the affine space.

	The foregoing discussion suggests the possibility of
re-thinking thermodynamics and information theory in a
novel mathematical framework \cite{qct-ternary}.
Both information theory and thermodynamics
are concerned with notions such as entropy, free energy and
relative entropy.  These concepts are introduced in Sec. \ref{sec:2}
under a single framework based on the Radon-Nikodym derivative,
as a random variable relating two different measures.
In its broadest context, we are able to capture
the essential mathematics used in the theory of
equilibrium and nonequilibrium thermodynamics.
This approach significantly enriches the scope of
``information theory'' \cite{cover-book}.  The
RN derivative should not be treated as an esoteric
mathematical concept: It is simply a powerful way to quantify
even infinitesimal changes in the probability distributions; it is the calculus for thinking of change in terms of chance \cite{yang-pre}.

In Sec. \ref{sec:3}, the notion of a temperature,
$T=\beta^{-1}$ is introduced through the canonical
probability distribution $Z^{-1}(\beta)e^{-\beta U(\omega)}$.
It has been shown recently that this Gibbsian
distribution has a much broader applications
than just thermal physics:  It is in fact a limit theorem of
a sequence of conditional probability densities under an
additive quasi-conservative observable \cite{cwq-19}.
The focus of this section is to show the centrality of
RN derivative in the theory of thermodynamics.  The RN derivative
is used to describe several results in physics
that includes the thermodynamic cycle, equation of
states, and the Jarzynski-Crooks equalities.

Next, in Sec. \ref{sec:4} we equip the space of probability measures
with an affine structure and show that
the canonical distribution with a random variable $U(\omega)$
and a parameter
$\beta$ becomes precisely an affine
line in the space of probability measures
when one particular measure $\mathbb{P}$ is chosen as
a reference point.  With this, the tangent space
becomes a linear vector space of random variables and it
provides a represenation for the space of probability
measures.   A series of results are obtained.
Readers who are more mathematically
inclined can skip the Sec. \ref{sec:3}, come directly
to Sec. \ref{sec:4}, and then go back to Sec. \ref{sec:3}
afterward.

	Sec. \ref{sec:5} contains some discussions.

	The presentation of the paper is not mathematically
rigorous.  The emphasis is on illustrating how the pure
mathematical concepts can be fittingly applied
in narrating this branch of physics.
More thorough treatments of the subject are
forthcoming \cite{thompson}.

\section{Entropy, relative entropy,
and a fundamental equation of information}
\label{sec:2}

\subsection{Information and entropy}

	Information theory owes (to a large extent) its existence
as a separate subject from the theories of probability and
statistics to a singular emphasis on the notion of
{\em entropy} as a quantitative measure of information.
It is important to point out at the outset that
{\em information} is a random variable, defined on a
probability space $(\Omega,\mathcal{F},\mathbb{P})$,
through a Radon-Nikodym derivative
$\frac{\rd \mathbb{P}}{\rd\mu}(\omega)$,
$\omega\in\Omega$, between two measures
$\mathbb{P}$ and $\mu$ that are absolutely
continuous w.r.t. each other
\cite{qian-pre01b,qian-zgkx,ye-qian-18}.
If the $\Omega\subseteq \mathbb{R}^n$ and $\mu$ is
the Lebesgue measure, then
\begin{equation}
           -\ln\left(\frac{\rd \mathbb{P}}{\rd \mu}(\omega)\right)
\label{self-info}
\end{equation}
is the {\em self-information} \cite{kolmogorov}, which
is a random variable and its expected value is the standard form
of Shannon entropy:
\begin{equation}
        S[\mathbb{P}] \triangleq  - \int_{\Omega} f(x)\ln f(x) \ \rd x,
\label{entropy}
\end{equation}
in which the Radon-Nikodym derivative is the probability
density function, $\frac{\rd \mathbb{P}}{\rd \mu} \equiv f(x)$.

In general, if $\mu$ is normalizable, then one has a maximum
entropy inequality $S[\mathbb{P}]\le \ln \mu(\Omega)<+\infty$.
Similarly, one has the free energy
\begin{equation}
         H[ \mathbb{P}\|\mu]
        \triangleq \int_{\Omega} \ln\left(\frac{\rd \mathbb{P}}{\rd \mu}(\omega)\right) \rd\mathbb{P}(\omega) \ge -\ln\mu(\Omega).
\label{freenergy}
\end{equation}
When $\mu$ is also a normalized probability measure
$\mathbb{P}'$, the $H[\mathbb{P}\|\mathbb{P}']$
is called the {\em relative entropy} or Kullback-Leibler (KL) divergence.
The minimum free energy inequality in (\ref{freenergy})
becomes the better known, but less interesting, $H[ \mathbb{P}\|\mathbb{P}']\ge 0$.

From now on, we will drop most references to the underlying space
$(\Omega, \mathcal{F})$.  Moreover, we will assume that $\Omega\subseteq \mathbb{R}^n$ with the usual $\sigma$-algebra and that $\mathbb{P}$ is absolutely continuous w.r.t. the Lebesgue measure.  These conditions are not strictly necessary, but they simplify the notation considerably in
illustrating our key ideas.

\subsection{Fundamental equation of information}

With the various forms of entropy introduced above and some straightforward
statistical logic, one naturally has the
following equation that involves  three measures: two
probabilistic and the Lebesgue.  In particular, let $\mathbb{P}_1$ and
$\mathbb{P}_2$ be two probability measures with density
functions $f_1(x)$ and $f_2(x)$ with respect to the Lebesgue measure:
\begin{eqnarray}
	\Delta S &=& S[\mathbb{P}_2] - S[\mathbb{P}_1] \ = \
       \int_{\mathbb{R}} f_1(x)\ln f_1(x) \rd x -
         \int_{\mathbb{R}} f_2(x)\ln f_2(x) \rd x
\nonumber\\[5pt]
	&=&  \underbrace{ \int_{\mathbb{R}} f_1(x)\ln\left(\frac{f_1(x)}{f_2(x)}\right) \rd x }_{\Delta S^{\text{(i)}}:\text{ entropy production}} +
       \underbrace{ \int_{\mathbb{R}}  \Big( f_2(x)-f_1(x)\Big)
               \Big(-\ln f_2(x)\Big) \rd x }_{\Delta S^{\text{(e)}}:\text{ entropy
                  exchange}}.
\label{fenet}
\end{eqnarray}
The entropy production $\Delta S^{\text{(i)}}$ is never negative, while
the entropy exchange $\Delta S^{\text{(e)}}$ has no definitive
sign.   If $f_2(x)$ is the unique invariant density of some
measure-preserving dynamics \cite{mackey}, then
$-\ln f_2(x)$ is customarilly referred to as the
``equilibrium energy function'', then
$\Delta S^{\text{(e)}}$ is the change in the
``mean energy'', which is related to ``heat''.

	Entropy and free energy in (\ref{entropy}) and (\ref{freenergy})
have their namesakes in the theory of statistical equilibrium
thermodynamics \cite{pauli}.  The Second Law, in terms
of entropy maximization or free energy minimization, has
its statistical basis precisely in the two inequalities associated
with $S$ and $H$.  The $\Delta S^{(\text{i})}$ term on the rhs of
(\ref{fenet}), however, is a nonequilibrium free energy
associated with a {\em nonequilibrium distribution}, either due to
a spontaneous fluctuation or a man-made
perturbation \cite{qian-pre01a}. In the theory of stochastic dynamics,
one uses a probability distribution $\rho(x,t)$ to represent the state
of a system; thus any $\rho$ that differs from the equilibrium
distribution is a nonequilibrium distribution.  In applications to
laboratory systems, the $\rho$ can only be obtained from
a data-based statistical approach.  This approach can rely on
either a time scale separation,
or a system of many independent and identically distributed
subsystems, or a fictitious ensemble.  Ideal gas theory and 
the Rouse model of polymers are two successful examples of the
second type \cite{qian-pre01a}.

	Eq. \ref{fenet} in fact has the form of the
{\em fundamental equation of nonequilibrium thermodynamics}.
It states that if $f_2(x)$ is uniform, then
$\Delta S = \Delta S^{\text{(i)}}\ge 0$; and if one identifies
$U(x)\triangleq-T\ln f_2(x)$, where $T$ is a positive constant,
then one can introduce
$F[\mathbb{P}] \triangleq \mathbb{E}^{\mathbb{P}}[U] - TS[\mathbb{P}]$, and
$\Delta F = T\Delta S^{\text{(i)}} \ge 0$.  Unifying the
various forms of the Second Law to
a single concept of entropy production was a key idea of
the Brussel school of thermodynamics \cite{prigogine}.\footnote{The second author
would like to acknowledge an enlightening discussion with
M. Esposito in the spring of 2011 at the Snogeholm Workshop
on Thermodynamics, Sweden.} See \cite{qian-ejpst,qkkb,ye-qian-18,qct-ternary},
and the references cited within, for the theory of entropy
production of Markov processes.

\subsection{Two results on relative entropy}

With regards to relative entropy, there are two results worth discussing.

	First, as the expected value of the logarithm of the
Radon-Nikodym derivative
$\xi\equiv\ln\left(\frac{\rd\mathbb{P}_1}{\rd\mathbb{P}_2}(\omega)\right)$, the relative entropy between two probability measures can be written as
\begin{equation}
        H\big[\mathbb{P}_1\|\mathbb{P}_2\big] = \int_{\mathbb{R}}
         f_1(x)\ln\left(\frac{f_1(x)}{f_2(x)}\right)\rd x
           = \mathbb{E}^{\mathbb{P}_1}\big[\xi(\omega)\big],
\end{equation}
with respective probability density
functions $f_1(x)=\frac{\rd\mathbb{P}_1(x)}{\rd x}$ and
$f_2(x)=\frac{\rd\mathbb{P}_2(x)}{\rd x}$.  The non-negativity of the $H[\mathbb{P}_1\|\mathbb{P}_2]$ can
actually be framed as a consequence of a stronger result, an
equality
\begin{equation}
     \mathbb{E}^{\mathbb{P}_1}\left[e^{-\xi(\omega)}\right] = 1,
\label{fpchs-eq}
\end{equation}
and an inequality for convex exponential function:
\begin{equation}
    \mathbb{E}^{\mathbb{P}_1}\big[\xi(\omega)\big]
 \ge -\ln \mathbb{E}^{\mathbb{P}_1}\left[e^{-\xi(\omega)}\right] = 0.
\end{equation}
Eq. \ref{fpchs-eq} implies that the Second Law and
entropy production could even be formulated through
equalities rather than inequalities.  Indeed, variations
of (\ref{fpchs-eq}) have found numerous applications in thermodynamics, such as Zwanzig's free energy perturbation
method \cite{zwanzig}, the Jarzynski-Crooks relation
\cite{jarzynski,crooks}, and the Hatano-Sasa
equality \cite{hatano-sasa}.

	Second, if the density $f_2$ contains an unknown parameter
$\theta$, then $f_2(x;\theta)$ is the likelihood
function for $\theta$.  In this case, with respect to the
change of measure,
\begin{eqnarray}
	\mathcal{I}_{\ell}(\theta) &\triangleq& -\mathbb{E}^{\mathbb{P}_2}\left[\left.
      \frac{\partial^\ell}{\partial\theta^\ell}\ln f_2(\omega;\theta)\right|\theta\right]
\nonumber\\
	&=& -\int_{\mathbb{R}} f_2(x;\theta) \frac{\partial^\ell}{\partial \theta^\ell}
               \ln f_2(x;\theta)\rd x
\nonumber\\
	&=& -\int_{\mathbb{R}} \left\{ \left(\frac{f_2(x;\theta)}{f_1(x)}\right)
      \frac{\partial^\ell}{\partial \theta^\ell}
               \ln\left(\frac{f_2(x;\theta)}{f_1(x)}\right)
      \right\} f_1(x)\rd x
\nonumber\\
	&=& \mathbb{E}^{\mathbb{P}_1} \left[ e^{-\xi(\omega)}
      \frac{\partial^\ell}{\partial \theta^\ell} \xi(\omega;\theta)
       \right].
\end{eqnarray}
$\mathcal{I}_0(\theta)$ is the Shannon entropy of $X_2(\theta)$,
$\mathcal{I}_1(\theta)\equiv 0$, and $\mathcal{I}_2(\theta)$ is
the {\em Fisher Information} for $X_2(\theta)$:
\begin{equation}
  \mathcal{I}_2(\theta) = \mathbb{E}\left[\left.
     \left( \frac{\partial}{\partial\theta}\ln f_2(X_2;\theta)\right)^2
       \right| \theta\right].
\end{equation}

\section{Canonical distribution and thermodynamics}
\label{sec:3}

	In many applications, stochastic dynamics exhibit
a separation of slow and fast time scales.  In mechanical
systems with sufficiently small friction, the dynamics are
organzied as fast Hamiltonian dynamics with slow energy
dissipation through heat.  The theory of thermodynamics arises
in this context when the mechanical motions of
point masses are described stochastically.   It can be shown then
that the probability distribution for the energy $E$ of a small
mechanical system in equilibrium with a large heat bath takes
a particularly canonical form
\begin{equation}
             p_E(y) = \frac{ \Omega^{(B)}(y)e^{-\beta y} }{Z(\beta)},
\label{canonicaldis}
\end{equation}
in which $\beta^{-1}=T$ is the temperature of the heat
bath \cite{pauli,cwq-19}.   In fact, if $\vx$ denotes random
variable in an appropriate state space and $U(x)$ is the mechanical
energy function, then one has distribution $f_{\vx}(x)\propto e^{-\beta U(x)}$, and
\begin{equation}
    p_E(y)\rd y =  \int_{y<U(x)\le y+\rd y} f_{\vx}(x)\rd x
                    =  \left( \frac{ \Omega^{(B)}(y) e^{-\beta y} }{Z(\beta)}
           \right)\rd y,
\end{equation}
in which
\begin{equation}
       \Omega^{(B)}(y) = \frac{1}{\rd y}
 \int_{y<U(x)\le y+\rd y} \rd x
          = \frac{\rd\Omega^{(G)}(y)}{\rd y}, \   \
         \Omega^{(G)}(y) = \int_{U(x)\le y} \rd x.
\label{B-G-entropy}
\end{equation}
$\ln\Omega^{(B)}$ and $\ln\Omega^{(G)}$ are called Boltzmann's entropy
and Gibbs' entropy in statistical mechanics \cite{g-vs-b-entropy}.
They are related via $\rd\Omega^{(G)}(y)=\Omega^{(B)}(y)\rd y$.
That is, $\Omega^{(G)}$ is a cumulative distribution function and $\Omega^{(B)}$ is its density function.

	Note that the expected value of any function of the
energy $U(x)$ (e.g., $g(U)$) is invariant under different representations
as a result of the rules of changes of variable for integration. For example, if $\vx$
is a state space representation and $E$ is the energy representation, then
\begin{eqnarray*}
  \int   g\big(U(x)\big) f_{\vx}(x)\rd x
	&=&  \int_{\mathbb{R}}  g(y)\left(\frac{e^{-\beta y}}{
             Z(\beta)}\right)  \Omega^{(B)}(y)\rd y
\\
	&=& \int_{\mathbb{R}} g(y)p_E(y) \rd y.
\end{eqnarray*}
In contrast, the thermodynamic entropy in statistical mechanics is
not invariant under different representations \cite{noyes}:
\begin{subequations}
\label{sm-entropy}
\begin{eqnarray}
  -\int p_{\vx}(x)\ln p_{\vx}(x)\rd x
	&=&  -\int_{\mathbb{R}} p_E(y)\ln
   \left(\frac{ p_E(y)}{\Omega^{(B)}(y)}\right)  \rd y
\\
	&\neq& -\int_{\mathbb{R}} p_E(y) \ln p_E(y) \rd y.
\end{eqnarray}
\end{subequations}
The rhs of (\ref{sm-entropy}a) is precisely
the negative free energy with non-normalized
$\Omega^{(G)}(y)$ as the reference measure (which has density
$\Omega^{(B)}(y)$).
The missing term from (\ref{sm-entropy}a) to (\ref{sm-entropy}b)
is contributed by the reference measure.  It is mean-internal-energy-like.
\begin{equation}
          \int_{\mathbb{R}} p_E(y)\Big(-\ln\Omega^{(B)}(y)\Big)\rd y.
\label{0014}
\end{equation}
We see that while $\ln \Omega^{(B)}(y)$ is widely considered
as an ``entropic'' term, it actually plays the role of an
energetic term in the energy represenation in
(\ref{sm-entropy}a).  In terms of this measure-theoretic framework, the distinction between entropy and energy is always relative.
This has long been understood in the work of J. G. Kirkwood
on the {\em potential of mean force}, which is itself
temperature dependent \cite{kirkwood-35}.

\subsection{Thermodynamics under a single temperature}

{\bf\em Equilibrium statistical thermodynamics.}
In terms of the canonical
distribution in (\ref{canonicaldis}), an equilibrium system
under a constant temperature $T=\beta^{-1}$ has its mechanical
energy distributed according to the canonical distribution
$p^{\text{eq}}(y)=Z^{-1}(\beta)\Omega(y)e^{-\beta y}$.  (We have dropped the superscript in $\Omega^{(B)}(y)$
to avoid cluttering.)  The {\em mean internal energy} associated
with the $p^{\text{eq}}(y)$ is then the expected value
\begin{subequations}
\label{eq13}
\begin{equation}
    \overline{U}(\beta) = \int_{\mathbb{R}} y\left(
      \frac{\Omega(y)e^{-\beta y}}{Z(\beta)}\right)
               \rd y = -\frac{\rd\ln Z(\beta)}{\rd \beta},
\end{equation}
which can be decomposed into an equilibrium {\em free energy}
and an {\em entropy}, $\overline{U}(\beta)=F^{\text{eq}}(\beta)
+\beta^{-1}S(\beta)$, where:
\begin{equation}
        \underbrace{F^{\text{eq}}(\beta) = -\beta^{-1}\ln Z(\beta)
     }_{\text{free energy}}   \text{ and }
    \underbrace{
   S(\beta)  = -\left(\frac{\rd F^{\text{eq}}(\beta)}{\rd \beta^{-1}}\right)
 }_{\text{ entropy}}.
\end{equation}
\end{subequations}
One can verify that the $S(\beta)$ is the same as
(\ref{sm-entropy}a), but not (\ref{sm-entropy}b).

{\bf\em Nonequilibrium statistical thermodynamics.}
For deep mathematical reasons that will become clear in
Sec. \ref{sec:4}, discussions of nonequilibrium systems should
begin in the full state space.  Intuitively, the canonical energy
representation $p^{\text{eq}}(E)$ based on a given energy
function $U(x)$ is a ``projection'' in the space of probability
measures that is nonholographic.

	Consider a system outside statistical equilibrium with a nonequilibrium
probability measure $\mu^{\text{neq}}$.  Suppose that this measure is absolutely continuous w.r.t. some other probability measure $\mathbb{P}$, with density
$\rho(x)=\frac{\rd\mu^{\text{neq}}}{\rd\mathbb{P}}(x)$.  The measure $\mu^{\text{neq}}$ possesses a nonequilibrium free energy
functional (a potential that can cause change) given by
\begin{subequations}
\label{Fneq0}
\begin{eqnarray}
      F^{\text{neq}}\big[\rho;\beta\big] &\triangleq&
          F^{\text{eq}}(\beta) + \beta^{-1}\int_{\Omega}
                  \rho(x)\ln\left(\frac{\rho(x)}{p^{\text{eq}}(x)}\right)
                  \rd x
\\
	&=&   \beta^{-1}\int_{\Omega}
                  \rho(x)\ln\left(\frac{\rho(x)}{e^{-\beta U(x)}}\
                    \right)\rd x.
\end{eqnarray}
\end{subequations}
One should recognize the fraction in (\ref{Fneq0}b) as a Radon-Nikodym derivative
of $\rho$ w.r.t. the non-normalized {\em canonical equilibrium measure} $e^{-\beta U(x)}$.  The minimum free energy inequality in
(\ref{freenergy}) takes the form
$F^{\text{neq}}\big[\rho;\beta\big] \ge F^{\text{eq}}(\beta)$
for any distribution $\rho$.  In fact,
$\beta\{F^{\text{neq}}\big[\rho;\beta\big]-F^{\text{eq}}(\beta)\}$
is the entropy production associated with the spontaneous
{\em relaxation process} of the
distribution $\rho$ tending to $p^{\text{eq}}$.

	The $ F^{\text{neq}}\big[\rho;\beta\big]$ also has another
expression:
\begin{equation}
     F^{\text{neq}}\big[\rho;\beta\big] =
 \underbrace{  \beta^{-1} \int_{\Omega}\rho(x)\ln\rho(x)\rd x
        }_{
   \text{neg-entropy}} +  \int_{\Omega}\rho(x)
            \underbrace{ \Big( -\beta^{-1}\ln p^{\text{eq}}(x)  +
           F^{\text{eq}}(\beta) \Big)
           }_{
   \text{internal energy of state $x$, $F^{\text{eq}}$ as reference }}
           \rd x.
\label{23}
\end{equation}
Eq. \ref{23} is very telling: The internal energy of a system in state $x$
is given in the second term with a fixed energy gauge ({\em i.e.}, the arbitrary
constant in the $U(x)$) according to the equilibirum $F^{\text{eq}}$, where $U(x)=F^{\text{eq}}(\beta)-\beta^{-1}\ln p^{\text{eq}}(x)$.
This fact implies that a change in the energy function from $U_1(x)$
to $U_2(x)$ necessarily involves a change of gauge.  Mechanical
work in classical thermodynamics can be understood as a
consequence of {\em gauge invariance}.
One particular $\beta$ defines an autonomous,
time-homogeneous stochastic dynamical system with a
unique $p^{\text{eq}}$.  All the energetic discussions
in such a system are with respect to the
equilibrium free energy $F^{\text{eq}}(\beta)$, which
fixes a choice for the energy gauge.  In the theory of probability,
the gauge invariance is achieved
through the notion of conditional probability and the law of
total probability.

\subsection{A clarification of Eq. \ref{Fneq0}}

A discussion on the meaning of the expression in (\ref{Fneq0}) is
in order.  To do that, let us only consider discrete
$x_k$, and the corresponding
\begin{equation}
   F^{\text{neq}}[\rho;\beta] = \sum_k
             \rho(x_k) \left[ \beta^{-1}
              \ln\left(\frac{\rho(x_k)}{p^{\text{eq}}(x_k)
          e^{-\beta F^{\text{eq}}(\beta)}
            }
      \right) \right].
\end{equation}
For a particular state $z$, if $\rho(x)=\delta_{x,z}$, then
$F^{\text{neq}}[\rho;\beta] =F^{\text{eq}}(\beta)-\beta^{-1}\ln p^{\text{eq}}(z)$,
which represents the traditional potential energy of the system in
the state $z$.  A question then naturally arises: Why is $F^{\text{neq}}[\rho;\beta]$ the average of
\begin{subequations}
\label{2-forms}
\begin{equation}
      \beta^{-1}\ln\left(\frac{\rho(x_k) }{p^{\text{eq}}(x_k)e^{-\beta F^{\text{eq}}}} \right),
\end{equation}
but not
\begin{equation}
      \beta^{-1}\ln\left(\frac{1}{p^{\text{eq}}(x_k)e^{-\beta F^{\text{eq}}}}
            \right)?
\end{equation}
\end{subequations}
Actually, (\ref{2-forms}b) is the potential energy for a deterministic initial state $x_k$.  It is natural, therefore, the average would be carried
out over the (\ref{2-forms}b) if the initial state of the system
were a {\em mixture of heterogeneous states} (mhs).  However,
if the initial state is a {\em stochastic fluctuating state} (sfs),
then the entropy of assimilation applies \cite{ben-naim} and
the $F^{\text{neq}}[\rho;\beta]$ in (\ref{Fneq0}a) is the average
carried out over the (\ref{2-forms}a).  The change from
mhs to sfs is analogous to a change from the Lagrangian to
the Eulerian representation in fluid mechanics; in
stochastic terms, the potential for an sfs to do work
is lower than an mhs \cite{parrondo}.

\subsection{Work, heat, and Jarzynski-Crooks' relation}

We now consider the case where the distribution $\rho(x)$ in (\ref{Fneq0})
arises from the equilibrium distribution $p^{\text{eq}}(x)$ as the consequence of a temperture change from
$T_a$ to $T_b$: $\rho(x) = Z^{-1}(\beta_a)e^{-\beta_a U(x)}$, and the $p^{\text{eq}}(x)=Z^{-1}(\beta_b)e^{-\beta_bU(x)}$.
Note that in the energy representation they can be written as
$\rho_E(y)=Z^{-1}(\beta_a)\Omega(y)e^{-\beta_a y}$ and
$p^{\text{eq}}_E(y)=Z^{-1}(\beta_b)\Omega(y)e^{-\beta_b y}$;
they share the same Gibbs entropy $\ln\Omega(y)$ determined
by $U(x)$ as in (\ref{B-G-entropy}).  Then
\begin{subequations}
\label{0024}
\begin{eqnarray}
	F^{\text{neq}}\big[\rho;\beta_b\big]-F^{\text{eq}}(\beta_b)
  &=& \beta_b^{-1}\int_{\Omega} \rho(x)\ln\left(
       \frac{\rho(x)}{p^{\text{eq}}(x)}\right) \rd x
\\
	&=& \beta_b^{-1}\int_{\mathbb{R}} \rho_E(y)\ln\left(
       \frac{\rho_E(y)}{p_E^{\text{eq}}(y)}\right) \rd y
\\
	&=&  \Big[\overline{U}(\beta_a)-\beta^{-1}_bS(\beta_a)\Big]-
   \Big[\overline{U}(\beta_b)-\beta^{-1}_bS(\beta_b)\Big].
\end{eqnarray}
\end{subequations}
The equation from (\ref{0024}a) to (\ref{0024}b) utilizes a key
property of a Radon-Nikodym derivative: {\em When it exists},
it is invariant under a change of measure.

	Eq. \ref{0024}c is not widely discussed, but it is a highly meaningful
result.  It contains the essence of Crooks' equality in
time-inhomogenous Markov processes \cite{crooks}.  It
implies that at the instant of switching from $T_a$ to
$T_b$, the system has internal energy $\overline{U}(\beta_a)$,
entropy $S(\beta_a)$, and nonequilibrium free energy
\begin{equation}
    F^{\text{neq}}[\rho;\beta] = \overline{U}(\beta_a)-T_b
                 S(\beta_a).
\label{the-eqn}
\end{equation}

	 Assuming that both
$\rho(x)$ and $p^{\text{eq}}(x)$ have the same $\Omega(y)$, Eq. \ref{0024} gives
the free energy change that is expected to be the maximum
reversible work that can be extracted.  We now explicitly consider
a change from $\rho(x)$ to $p^{\text{eq}}(x)$ that involves
changing the mechanical energy function from $U_1(x)$
to $U_2(x)$.  Even though the
corresponding canonical energy distributions are $\rho_E(y)=Z^{-1}_1(\beta_a)\Omega_1(y)e^{-\beta_ay}$ and $p_E^{\text{eq}}(y)=Z^{-1}_2(\beta_b)\Omega_2(y)e^{-\beta_by}$,
these RN derivative $\frac{\rd\rho_E}{\rd p^{\text{eq}}}(\omega)$
can be infinity!  Thus in this case one has to
start with the full distributions on the state space:
\begin{eqnarray}
 \beta_b^{-1}\int_{\Omega} \rho(x)\ln\left(
       \frac{\rho(x)}{p^{\text{eq}}(x)}\right) \rd x
  &=& \Big[\overline{U}_1(\beta_a)-\overline{U}_2(\beta_b)\Big] -\beta^{-1}_b\Big[S_1(\beta_a) -S_2(\beta_b)\Big]
\nonumber\\
	&+&
    \int_{\Omega}  \rho(x) \big[ U_2(x)- U_1(x)\big]
          \rd x.
\label{0021}
\end{eqnarray}
The last term in (\ref{0021}) is identified as the irreversible
work associated with the isothermal relaxation process
with mechanical change from $U_1(x)$ to $U_2(x)$,
\begin{equation}
\label{def-work}
 \overline{\mathcal{W}}_{12}(\beta_a) =  \int_{\Omega}  \rho(x) \mathcal{W}_{12}(x)
          \rd x,
\end{equation}
in which $ \mathcal{W}_{12}(x)$ should be considered as
the logarithm of the Radon-Nikodym derivative between
two non-normalized measures
\begin{equation}
             \mathcal{W}_{12}(x) =  \beta_a^{-1}\ln
     \left(\frac{ e^{-\beta_aU_1(x)}}{ e^{-\beta_aU_2(x)} }\right)
     =  \beta_b^{-1}\ln
     \left(\frac{ e^{-\beta_bU_1(x)}}{ e^{-\beta_bU_2(x)} }\right).
\label{work}
\end{equation}
$\mathcal{W}_{12}(x)$ is actually not a function of $\beta$;
work done in an isothermal process is independent of
the temperature.  In the canonical energy representation
of $U_1(x)$, then,
\begin{eqnarray}
     \overline{\mathcal{W}}_{12}(\beta_a) &=&
  \int_{\Omega}  \rho(\vx) \big[  U_2(x)-U_1(x)\big]
          \rd x
\nonumber\\
	&=& \int_{\mathbb{R}}  \left( \frac{ \Omega_1(y)e^{-\beta_ay} }{Z_1(\beta_a)}\right)  \left\{ \frac{\displaystyle\int_{y<U_1(x)\le y+\rd h }
          U_2(x)\rd x }{\displaystyle \int_{y<U_1(x)\le y+\rd h } \rd x }  - y \right\} \rd y.
\label{e25}
\end{eqnarray}
The first term inside $\{\cdots\}$ is a conditional expectation:
$\mathbb{E}^{^{\text{eq}}}\big[ U_2(\vx)\big| U_1(\vx)=y\big]$,
where $\mathbb{E}^{^{\text{eq}}}$ is the expectation in
terms of the equilibrium measure $p^{\text{eq}}(x)$.

The transfered irreversible heat is
\begin{equation}
   \mathcal{Q}(\beta_b) \triangleq
  \beta^{-1}_b \left\{ S_1(\beta_a) -S_2(\beta_b) +
   \int_{\Omega} \rho(x)\ln\left(
       \frac{\rho(x)}{p^{\text{eq}}(x)}\right) \rd x  \right\}.
\label{heat}
\end{equation}
Then the relation
\begin{equation}
    S_2(\beta_b)-S_1(\beta_a)+\frac{ \mathcal{Q}(\beta_b) }{T_b}
       = \Delta S^{\text{(i)}} = \int_{\Omega} \rho(x)\ln\left(
       \frac{\rho(x)}{p^{\text{eq}}(x)}\right) \rd x \ge 0
\end{equation}
is known as the Clausius inequality in thermodynamics.  The equality
is a special case of the fundamental equation of nonequilibrium
thermodynamics.

	Concerning the work $\mathcal{W}_{12}(x)$ in (\ref{work}),
we have Jarzynski-Crooks' relation \cite{jarzynski,crooks}:
\begin{eqnarray}
	 \int_{\Omega}  \left( \frac{ e^{-\beta_aU_1(x)} }{Z_1(\beta_a)}\right)  e^{-\beta_a\mathcal{W}_{12}(x)}
         \rd x =   \int_{\Omega}   \frac{e^{-\beta_aU_2(x)} }{Z_1(\beta_a)}\rd x = \frac{Z_2(\beta_a)}{Z_1(\beta_a)}.
\label{eq0025}
\end{eqnarray}
Note that the work is performed under $\beta_b$, but the
rhs of (\ref{eq0025}) is evalued at $\beta_a$.
The original Jarzynski-Crooks' equality emphasized
path-wise average over a stochastic trajectory, but
Eq. \ref{eq0025} is an ensemble average over a single
step, which can be generalied to many different other
forms \cite{quanhaitao}.

{\bf\em The concept of exergy.}
In Eq. \ref{the-eqn}, equilibrium internal energy and
entropy under temperature $T_a$, $\overline{U}(T_a)$
and $S(T_a)$ are assembled with temperature
$T_b\neq T_a$ to form a nonequilibrium free energy
$F^{\text{neq}}=\overline{U}(T_a)-T_bS(T_a)$, which
plays a central role in our analysis of
canonical systems.  This quantity has been extensively
discussed in the literature on thermodynamics: {\em Exergy} of a system is ``the maximum fraction of an energy form which can be transformed into work.''
The remaining part is the waste heat \cite{honerkamp-book}.
After a system reaches equilibrium with its surrounding,
its exergy is zero.  Therefore, the concept of exergy
epitomizes a nonequilibrium quantity \cite{chen-wu-sun}.
Its identification to the entropy production in Eq. \ref{0024} implies
its importance in information energetics.  Even though the term
``exergy'' was coined as late as in 1956, the idea had been
already in the work of Gibbs.

{\bf\em Mechanical work of an ideal gas.}
For an ideal gas with total mechanical energy
$U(x) = U_{p}(x_1) +U_{k}(x_2)$, where
$U_p$ and $U_k$ are potential and kinetic energy functions,
and $\vx_1$ and $\vx_2$ are position and momentum
state variables,
\begin{equation}
    U(x) = \sum_{i=1}^N \left\{ \frac{x_{2,i}^2}{2m_i}
                 + H_V\big(x_{1,i}\big) \right\},
\end{equation}
in which $H_V(z)=0$ when $0<z<V$ and $H_V(z)=+\infty$ when
$z\le 0$ or $z\ge V$.  The $V$ represents the ``volume'' of a box
containing the ideal gas.  Then
\begin{equation}
     \Omega(E,V) = \frac{ V^N}{\rd E}
\int_{E<U_{k}(x_2)\le E+\rd E}
       \rd x_2 = V^N \tilde{\Omega}(E,N),
\end{equation}
in which the $\tilde{\Omega}$ is independent of $V$.
Therefore, the mechanical work associated with a change in $V_1=V\to V_2=V+\Delta V$ is given by
\begin{equation}
   \beta^{-1}\ln\left(\frac{ \Omega(E,V_2)}{ \Omega(E,V_1) } \right)
     = NT\ln\left(\frac{V+\Delta V}{V}\right)
     = \frac{NT\Delta V}{V} = \hat{p}\Delta V,
\end{equation}
where $\hat{p}=Nk_BT/V$ is the pressure of an ideal gas.
(We have set Boltzmann's constant $k_B\equiv 1$ throughout the present
paper.)

\subsection{Application to heat engines and thermodynamic cycles}

{\bf\em Carnot cycle.}
Applying Eqs. \ref{work} and \ref{heat} twice for
{\em thermomechanical} ({\em i.e.}, temperature and mechanical) changes
from $\{T_a,U_1\}$ to $\{T_b,U_2\}$
and from $\{T_b,U_2\}$ back to $\{T_a,U_1\}$, we
derive the celebrated Carnot efficiency for a heat engine.
For each of the processes descibed in the left column below, the
energetic status of the system is shown in the right column:
\begin{subequations}
\label{Cc}
\begin{eqnarray}
   \text{adiabatic switching $\{T_a,U_1\}\to\{T_b,U_1\}$:}  &&
           F_1^{\text{neq}}(T_b) = \overline{U}_1(T_a)-T_bS_1(T_a),
\\
		 \text{isothermal relaxation $\{T_b,U_1\}\to\{T_b,U_2\}$:}
	 	&&   \overline{U}_1(T_a)-\overline{U}_2(T_b)
            =  \mathcal{Q}_{12}(T_b) - \overline{\mathcal{W}}_{12},\nonumber\\
\\
	 \text{equilibrium under $T_b$:} &&
          F^{\text{eq}}_2(T_b) = \overline{U}_2(T_b)-T_bS_2(T_b),
\\
   \text{adiabatic switching $\{T_b,U_2\}\to\{T_a,U_2\}$:} &&
           F_2^{\text{neq}}(T_a) = \overline{U}_2(T_b)-T_aS_2(T_b),
\\
		 \text{isothermal relaxation $\{T_a,U_2\}\to\{T_a,U_1\}$:}
	 	&&   \overline{U}_2(T_b)-\overline{U}_1(T_a)
            =  \mathcal{Q}_{21}(T_a) - \overline{\mathcal{W}}_{21},\nonumber\\
\\
	 \text{equilibrium under $T_a$:} &&
          F^{\text{eq}}_1(T_a) = \overline{U}_1(T_a)-T_aS_1(T_a).
\end{eqnarray}
\end{subequations}
In (\ref{Cc}f), the system is returned to
the equilibrium state under $T_a$.  Without loss of generality,
let $T_a>T_b$.  In the ideal Carnot cycle, one assumes that the
processes of switching the temperatures are adiabatic without
free energy dissipation.  That is, the $F^{\text{neq}}_1(T_b)$
in (\ref{Cc}a) is strictly equal to $F^{\text{eq}}_1(T_a)$ in
(\ref{Cc}f), with a reversible change of gauge reference,
and similarly the $F^{\text{neq}}_2(T_a)$ in (\ref{Cc}d) is
strictly equal to $F^{\text{eq}}_2(T_b)$ in
(\ref{Cc}c).  In the two processes of isothermal relaxation, irreversible heat $\mathcal{Q}_{12}(T_b)=T_b\{S_1(T_a)-S_2(T_b)+\Delta S^{\text{(i)}}_{12}\}$ and $\mathcal{Q}_{21}(T_a)=T_a\{S_2(T_b)-S_1(T_a)+\Delta S^{\text{(i)}}_{21}\}$ each contain an entropy production term,
\begin{equation}
\Delta S^{\text{(i)}}_{jk}=\int_{\mathbb{R}} p^{\text{eq}}_j(y)\ln\left(
       \frac{p^{\text{eq}}_j(y)}{p^{\text{eq}}_k(y)}\right) \rd y\ge 0.
\end{equation}
In a Carnot cycle
with {\em quasi-static} processes, they are assumed to be zero.
Then, the total work done by the system
over the cycle is
\begin{eqnarray}
    W &=& -\Big(\overline{\mathcal{W}}_{12} +
      \overline{\mathcal{W}}_{21}\Big) \ = \ -\mathcal{Q}_{12}(T_b)-
         \mathcal{Q}_{21}(T_a)
\nonumber\\
	&=&  T_b \left\{  S_2-S_1 -
   \int_{\mathbb{R}}p^{\text{eq}}_1\ln\left(
       \frac{p^{\text{eq}}_1}{p^{\text{eq}}_2}\right) \rd y  \right\}+
T_a \left\{  S_1-S_2 -
   \int_{\mathbb{R}} p^{\text{eq}}_2\ln\left(
       \frac{p^{\text{eq}}_2}{p^{\text{eq}}_1}\right) \rd y  \right\}
\nonumber\\
	&\le&  T_b \Big[ S_2(T_b)-S_1(T_a)\Big] + T_a\Big[ S_1(T_a)-S_2(T_b) \Big],
\label{eq32}
\end{eqnarray}
in which the reversible heat being absorbed at $T_a$ is
$Q_h=T_a[S_1(T_a)-S_2(T_b)]>0$, and the heat being expelled
at $T_b$ is $Q_l=T_b[S_2(T_b)-S_1(T_a)]<0$.
Thus the Carnot (first-law) efficiency
\begin{equation}
    \eta_{Carnot} = \frac{W}{Q_h}
    \le 1 - \frac{T_b}{T_a}.
\label{eq0031}
\end{equation}
On the other hand, since the rhs of (\ref{eq32}) is the maximum
possible work, the second-law, exergy efficiency
\begin{equation}
     \eta_{exergy} = \frac{W}{(T_a-T_b)[S_1(T_a)-S_2(T_b)]}
                         = \frac{W}{Q_h\left(1-\frac{T_b}{T_a}\right)}
                         \le 1.
\label{eq0034}
\end{equation}

{\bf\em Stirling cycle.}
There are many different realizations of heat engines in terms of
thermodynamic cycles. We now consider the Stirling cycle below.
\begin{subequations}
\label{Sc}
\begin{eqnarray}
   \text{isothermal working $\{T_a,U_1\}\to\{T_a,U_2\}$:} &&
           \overline{U}_1(T_a)-\overline{U}_2(T_a)
            =  \overline{\mathcal{Q}}_{12}(T_a) - \overline{\mathcal{W}}_{12},\nonumber\\
\\
		 \text{isochoric cooling $\{T_a,U_2\}\to\{T_b,U_2\}$:}
	 	&&    \overline{U}_2(T_a)-\overline{U}_2(T_b)
            =  \mathcal{Q}_{2}(T_a,T_b),
\\
	 \text{equilibrium under $\{T_b,U_2\}$:} &&
          F^{\text{eq}}_2(T_b) = \overline{U}_2(T_b)-T_bS_2(T_b),
\\
   \text{isothermal working $\{T_b,U_2\}\to\{T_b,U_1\}$:} &&
           \overline{U}_2(T_b)-\overline{U}_1(T_b)
            =  \overline{\mathcal{Q}}_{21}(T_b) - \overline{\mathcal{W}}_{21},\nonumber\\
\\
		 \text{isochoric heating $\{T_b,U_1\}\to\{T_a,U_1\}$:}
	 	&&    \overline{U}_1(T_b)-\overline{U}_1(T_a)
            =  \mathcal{Q}_{1}(T_b,T_a),
\\
	 \text{equilibrium under $\{T_a,U_1\}$:} &&
          F^{\text{eq}}_1(T_a) = \overline{U}_1(T_a)-T_aS_1(T_a).
\end{eqnarray}
\end{subequations}
After two isothermal processes in (\ref{Sc}a), (\ref{Sc}d), the system is still in the equilibrium states with free energy $F^{\text{eq}}_2(T_a) = \overline{U}_2(T_a)-T_aS_2(T_a)$ and $F^{\text{eq}}_1(T_b) = \overline{U}_1(T_b)-T_bS_1(T_b)$ respectively. Notice the difference between the equilibrium free energy above and the non-equilibrium free energy functions $F^{\text{neq}}_2(T_a)$ and $F^{\text{neq}}_1(T_b)$ defined in (\ref{Cc}a) and (\ref{Cc}d). The irreversible heats for the two isothermal processes are
\begin{eqnarray}
    \overline{\mathcal{Q}}_{12}(T_a) &=& T_a\left[S_1(T_a)-S_2(T_a)+\int_{\Omega}\rho_1(x;T_a)\ln\left(\frac{\rho_1(x;T_a)}{\rho_2(x;T_a)}\right) \rd x\right],
\\
  \overline{\mathcal{Q}}_{21}(T_b) &=& T_b\left[S_2(T_b)-S_1(T_b)+\int_{\Omega}\rho_2(x;T_b)\ln\left(\frac{\rho_2(x;T_b)}{\rho_1(x;T_b)}\right) \rd x\right].
\end{eqnarray}
Meanwhile, those for the isochoric cooling and heating processes are \begin{eqnarray}
    \mathcal{Q}_{2}(T_a,T_b) &=&T_b\left[S_2(T_a)-S_2(T_b)+\int_{\Omega}\rho_2(x;T_a)\ln\left(\frac{\rho_2(x;T_a)}{\rho_2(x;T_b)}\right) \rd x\right],
\\
\mathcal{Q}_{1}(T_b,T_a) &=& T_a\left[S_1(T_b)-S_1(T_a)+\int_{\Omega}\rho_1(x;T_b)\ln\left(\frac{\rho_1(x;T_b)}{\rho_1(x;T_a)}\right) \rd x\right].
\end{eqnarray}
Summarizing the whole heat cycle, we find that
\begin{eqnarray}
 W &=& -\Big(\overline{\mathcal{W}}_{12} +
      \overline{\mathcal{W}}_{21}\Big) \ = \ -\overline{\mathcal{Q}}_{12}(T_a)-\mathcal{Q}_{2}(T_a,T_b)-
         \overline{\mathcal{Q}}_{21}(T_b)-\mathcal{Q}_{1}(T_b,T_a)\nonumber\\
 &\leq& (T_a-T_b)\big[S_2(T_a)-S_1(T_b)\big].
\label{stirling-eff}
\end{eqnarray}
This will lead to the same conclusions on the first-law and second-law efficiency for the Stirling cycle.

{\bf\em Realization of a reversible cycle.}
The Carnot cycle and Stirling cycle considered above are not truly reversible, once $U_1\neq U_2$ or $T_a\neq T_b$. To achieve the theoretical maximal efficiency, we need to construct a reversible heat cycle through a series of quasi-static processes, each of which involves only an infinitesimal change in either $U$ or $T$.
Taking the Stirling cycle as an example. In the first isothermal working step, we insert $N-1$ intermediate states between $\{T_a,U_1\}$ and $\{T_a,U_2\}$, that are $\{T_a,U_1+\triangle U\}, \{T_a,U_1+2\triangle U\}, \cdots, \{T_a,U_1+(N-1)\triangle U\}$ with $\triangle U=(U_2-U_1)/N$. In the limit of $N\rightarrow\infty$, $\triangle U\rightarrow0$, which means each transition between two adjacent states can be treated as a quasi-static process. Therefore, the whole step between $\{T_a,U_1\}$ and $\{T_a,U_2\}$ becomes reversible with the help of those intermediate states. Applying similar procedure to other three steps, we will achieve a true thermodynamically reversible Stirling cycle by requiring an infinitesimal change in either $U$ or $T$ for each sub-step.

\subsection{Work as a conditional expectation in energy representation}

	Consider once again two distributions $\rho(x)$ and $p^{\text{eq}}(x)$ with respective energy representations,
$\rho_E(y)=Z^{-1}_1(\beta_a)\Omega_1(y)e^{-\beta_ay}$ and $p_E^{\text{eq}}(y)=Z^{-1}_2(\beta_b)\Omega_2(y)e^{-\beta_by}$.
The key thermodynamic quantity that arises in (\ref{0021}),
the irreversible work, can not be expressed in terms
of the six quantities: $\Omega_1(y),Z_1(\beta)$,
$\Omega_2(y),Z_2(\beta)$, and $\beta_a,\beta_b$.
We note that
\begin{equation}
     \int_{\Omega}  \rho(x) \big[ U_2(x)- U_1(x)\big]
          \rd x
	=  \int_{\mathbb{R}} \left(\frac{ \Omega_1(y) e^{-\beta_ay}}{
       Z_1(\beta_a)}\right)\Big\{  \overline{U}_{2|U_1=y}
             - y\Big\}\ \rd y,
\label{eq0036}
\end{equation}
in which
\begin{eqnarray}
  \overline{U}_{2|U_1=y} =
       \frac{\displaystyle \int_{y<U_1(x)\le y+\rd h}
                  U_2(x)
                  \rd x }{\displaystyle  \int_{y<U_1(x)\le y+\rd h}
                  \rd x },
\label{eq0037}
\end{eqnarray}
is a conditional expectation of $U_2(x)$ given $U_1(x)=y$.
The energy functions $U_1(x)$ and $U_2(x)$ are only two
observables on the probability space and they certainly do not provide
a full description of the probability space.  Actually,
knowing the canonical energy distributions $\rho_E(y)$ and
$p^{\text{eq}}_E(y)$ is not equivalent to knowing their joint probability distribution; the missing information on their correlation
is captured precisely in (\ref{eq0037}).

	The lhs of (\ref{eq0036})  can also be expressed as
\begin{eqnarray}
 && \int_{\Omega}  \rho(x) \Big[ U_2(x)- U_1(x)\Big]
          \rd x
\nonumber\\
  &=&  \frac{1}{\beta_a}\left[ \ln\frac{Z_1(\beta_a)}{Z_2(\beta_a)} + \int_{\Omega}  \rho(x) \ln
           \left\{ \frac{ e^{ -\beta_a U_1(x)} Z_2(\beta_a)}{Z_1(\beta_a)
  e^{-\beta_a U_2(x)} } \right\} \rd x \right].
\end{eqnarray}
The term inside $\{\cdots\}$ indeed can be understood
as a Radon-Nikodym derivative between the two
probability measures, which is well-defined on the
entire $\sigma$-algebra $\mathcal{F}$ as well as
the restricted joint $\sigma$-algebra
$\mathcal{F}_{U_1,U_2}$.  However, it is singular
on the further restricted $\sigma$-algebra $\mathcal{F}_{U_1}$
or $\mathcal{F}_{U_2}$.

\subsection{The role and consequence of determinism}
\label{sec:3.6}

Consider a sequence of measures $\mu_{\epsilon}$ and
two real-valued continuous
random variables $\vx(\omega)$ and
$\vy(\omega)$,
with corresponding probability density functions
$p_{\epsilon}(x)$ and $q_{\epsilon}(x)$.
Their relative entropy is then
\begin{equation}
        H\big[\vx\|\vy;\mu_{\epsilon}\big]
      = \int_{\Omega}  p_{\epsilon}(x)
 \ln \left(\frac{ p_{\epsilon}(x)}{
           q_{\epsilon}(x)}\right) \rd x.
\label{034}
\end{equation}
If the sequence of measures $\mu_{\epsilon}$ tends to a
singleton with corresponding $p_{\epsilon}(x)\to \delta(x-z)$ and $q_{\epsilon}(x)\to \delta(x-y^*)$ as $\epsilon\to 0$,
we call the limit {\em deterministic}.

	It can be shown under rather weak conditions, or more
properly through the theory of large deviations, that
as $\epsilon\to 0$ the $p_{\epsilon}(x)$ and $q_{\epsilon}(x)$
have asymptotic forms
\begin{equation}
            \ln p_{\epsilon}(x) = -\frac{\varphi_p(x)}{\epsilon}
                + O(\ln\epsilon),  \
            \ln q_{\epsilon}(x) = -\frac{\varphi_q(x)}{\epsilon}
                + O(\ln\epsilon),
\end{equation}
in which $\varphi_p(z)=\varphi_q(y^*)=0$.
This asymptotic relation is known as the large deviations 
principle in the theory of probability \cite{touchette}. 
Therefore,
\begin{equation}
        \ln H\big[\vx\|\vy; \mu_{\epsilon}\big]
      \sim  \frac{\varphi_q(z)}{\epsilon}
                + O(\ln\epsilon),
\end{equation}
as $\vx\to z$. Even though $\vy\to y^*$, the relative entropy
in (\ref{034}) provides the $\varphi_q$ as a function of
$z$ fully supported on $\mathbb{R}^n$.
If the $q_{\epsilon}$ is an invariant measure
of a stochastic dynamical system, then the $\varphi_q(z)$ is
thought of as a ``deterministic energy function'', which
can be obtained as the asymptotic limit of determinism.
The normalization of $e^{-\varphi_q(x)/\epsilon}$,
however, is lost in the $\ln\epsilon$-order term.  This
corresponds to a certain gauge freedom.

	A combination of the determinism with the canonical
distribution immediately yields a key relationship
that is well known in thermodynamics.
Specifically, if the probability density function
\begin{equation}
     \frac{\Omega^{(B)}(E)e^{-\beta E}}{Z(\beta)}
       =  \frac{e^{-\beta E+\ln\Omega^{(B)}(E)}}{Z(\beta)}
    \to \delta\big(E-E^*\big),
\end{equation}
in an asymptotic limit, then one has the {\em equation of state}
\begin{eqnarray}
		\left[ \frac{\rd}{\rd E} \Big(\beta E-\ln\Omega^{(B)}(E) \Big) \right]_{E=E^*} = 0.
\label{eq39}
\end{eqnarray}
A system in macroscopic thermodynamic equilibrium possesses
one less degree of freedom \cite{pauli}.  Eq. \ref{eq39} implies
\begin{equation}
     \beta = \frac{\rd\ln\Omega^{(B)}(E^*)}{\rd E}
         =\frac{\frac{\rd}{\rd E}\Omega^{(B)}(E^*)}{\Omega^{(B)}(E^*)},
\end{equation}
in which
\begin{eqnarray}
  \Omega^{(B)}(E) &=& \frac{1}{\rd E}\int_{E<U(x)\le E+\rd E}
          \rd x
	\ = \  \oint_{U(x)=E}
          \frac{\rd\Sigma\cdot \hat{\bf{n}} }{\|\nabla U(x)\|}
\nonumber\\
    &=&  \int_{U(x)\le E}
          \nabla\cdot\left( \frac{\nabla U(x) }{\|\nabla U(x)\|^2}\right)
         \rd x,
\\[6pt]
	\frac{\rd\Omega^{(B)}(E)}{\rd E} &=& \frac{1}{\rd E}
      \int_{E<U(x)\le E+\rd E}
   \nabla\cdot\left( \frac{\nabla U(x) }{\|\nabla U(x)\|^2}\right) \rd x
\nonumber\\
	&=&  \oint_{U(x)\le E}
         \nabla\cdot\left( \frac{\nabla U(x) }{\|\nabla U(x)\|^2}\right)
          \frac{\rd\Sigma\cdot \hat{\bf{n}} }{\|\nabla U(x)\|}.
\end{eqnarray}
Therefore,
\begin{equation}
\frac{\rd\ln\Omega^{(B)}(E)}{\rd E} = \frac{\displaystyle  \oint_{U(x)= E}
         \nabla\cdot\left( \frac{\nabla U(x) }{\|\nabla U(x)\|^2}\right)
          \frac{\rd\Sigma\cdot \hat{\bf{n}} }{\|\nabla U(x)\|}}{
         \displaystyle  \oint_{U(x)=E}
          \frac{\rd\Sigma\cdot \hat{\bf{n}} }{\|\nabla U(x)\|} }.
\label{00047}
\end{equation}
That is, the equilibrium $\beta$ is the
average of
\begin{equation}
    \nabla\cdot\left(\frac{ \nabla U}{\|\nabla U\|^2}\right)
         = \frac{\|\nabla U\|\nabla^2 U-
            2\nabla U\cdot\nabla\|\nabla U\| }{\|\nabla U\|^3},
\end{equation}
on the level-surface $\{x:U(x)=E^*\}$.  For a given
energy function $U(x)$, or an observable \cite{cwq-19},
Eq. \ref{00047}, which generalizes the virial theorem in
classical mechanics, provides the function $\beta(E)$.

\section{The space of probability measures}
\label{sec:4}

\subsection{Affine structure, canonical distribution and its energy
representation}

	We will now give a brief, non-rigorous introduction to the theory
developed in \cite{thompson}.
Let $\mathcal{M}$ be the set of all probability
measures on $(\Omega,\mathcal{F})$ that are absolutely continuous w.r.t. some probability measure $\mathbb{P}$ (and therefore absolutely continuous w.r.t. each other) and let $\mathcal{V}$
be an appropriate set of real-valued functions on $\Omega$.  (Note that any choice of $\mathbb{P}$ in $\mathcal{M}$ would do; one only cares that all measures in $\mathcal{M}$ are absolutely continuous w.r.t each other.)  One now
defines $\oplus$:
$\mathcal{M}\times\mathcal{V}\to\mathcal{M}$ such
that
\begin{equation}
    \big(\mu\oplus g\big)(A) = \frac{\displaystyle \int_{A} e^g \rd\mu}{
         \displaystyle \int_{\Omega} e^g\rd\mu},
\end{equation}
for any $A\in\mathcal{F}$.  Assuming the denominator is finite (which requires some assumptions on $\mathcal{V}$), the positivity of $e^{g}$ implies that $(\mu\oplus g)$ is also absolutely continuous w.r.t. $\mathbb{P}$.
Since $(\mu\oplus g)(\Omega)=1$, it is a probability measure.  These two facts mean that $(\mu\oplus g)\in\mathcal{M}$, so the operation $\oplus$ is well-defined.  Note that $\mu\oplus g = \mu\oplus (g + c)$ for any constant $c$, so this addition is not actually one-to-one.  We can remedy this issue by restricting $\mathcal{V}$ to functions that sum to zero, or we can replace each function with an equivalence class of functions that differ by a constant.  
One can then show that $(\mathcal{M},\mathcal{V},\oplus)$ is an
affine structure on $\mathcal{M}$ \cite{jeangallier}, \cite{thompson}.
If one chooses a particular measure
$\mathbb{P}\in\mathcal{M}$ as the origin, then
any other measure $\mu\in\mathcal{M}$ will have a Radon-Nikodym derivative
$\frac{\rd\mu}{\rd\mathbb{P}}(\omega)$,
and $\mu = (\mathbb{P}\oplus g)$
where $g=\ln\big(\frac{\rd\mu}{\rd\mathbb{P}}\big)$.

	Let $J\subseteq\mathbb{R}$ be an interval and
$U\in\mathcal{V}$.  The function $p$: $J\to\mathcal{M}$
such that $p(\beta)=\mathbb{P}\oplus(-\beta U)$ is an affine straight line.   More explicitly,
we have the family of probability densities
\begin{equation}
    \frac{e^{-\beta U(\omega)}}{Z(\beta)}\mathbb{P}(\rd\omega),
\label{51}
\end{equation}
where $Z(\beta)$ is the normalization factor.

	In Kolmogorov's theory, the real-valued
function $U(\omega)$, when thought of as a random variable, has its own
probability density function w.r.t. the Lebesgue measure:
\begin{equation}
    \mathbb{P}\Big\{ y<U(\omega)\le y+\rd h\Big\}
        = \frac{\Omega_U(y)}{Z(\beta)}e^{-\beta y}\rd h,
\label{52}
\end{equation}
in which $\ln \Omega_U(y)$ is the Gibbs entropy associated
with function $U(\omega)$, defined in Eq. \ref{B-G-entropy}:
\begin{equation}
            \Omega_U(y) = \frac{1}{\rd h}\int_{y<U(\omega)\le
   y+\rd h} \mathbb{P}(\rd\omega).
\end{equation}
The relation between the distributions in (\ref{51}) and
(\ref{52}) establishes a map between the observables in
the tangent space $\mathcal{V}$ of $\mathcal{M}$ and
the standard probability density functions.  (This is
analogous to the dual relation between the Koopman operator
on the space of observables and the Perron-Frobenius operator
on the space of densities in dynamical systems theory.)
We call (\ref{51}) the {\em canonical representation} for
the space of probability measures (SoPMs), and
(\ref{52}) its {\em energy represenation}.  Note that the energy representation of a given probability measure is not unique.  The choice of $U$ depends on both $\mathbb{P}$ and $\beta$.  

{\bf\em A pair of observables.}
We now discuss the notions of joint, marginal, and
conditional probability in terms of the canonical representation
in $\mathcal{V}$, with a fixed ``origin''
$\mathbb{P}$, which
should be thought of as the $\mathbb{P}$ in the probability space
$(\Omega,\mathcal{F},\mathbb{P})$, {\em \`{a} la}
Kolmogorov.   The SoPMs $\mathcal{M}$ then
is represented by observables $U(\omega)\in\mathcal{V}$,
the tangent space of $\mathcal{M}$.

	Consider two observables $\big(U_1(\omega),U_2(\omega)\big)$, where
$U_2\neq aU_1+b$.  The corresponding ``flat plane'' can be parametrized as
\begin{equation}
            \frac{  e^{-\beta_a U_1(\omega)-\beta_b U_2(\omega)}
       }{Z_{1,2}(\beta_a,\beta_b)}\mathbb{P}(\rd\omega), \    (\beta_a,\beta_b)\in\mathbb{R}^2.
\label{2d}
\end{equation}
We note that each observable induced a restricted $\sigma$-algebra on $\mathbb{R}$:
$\mathcal{F}_{U_1}$ and $\mathcal{F}_{U_2}$ respectively, and the
joint observable induces $\mathcal{F}_{U_1,U_2}=\sigma(\mathcal{F}_{U_1}\cup\mathcal{F}_{U_2})$.  With respect to  $\mathcal{F}_{U_1,U_2}$, the distribution
in (\ref{2d}) can expressed on as:
\begin{subequations}
\begin{equation}
    \frac{\Omega_{1,2}(y_1,y_2) }{Z_{1,2}(\beta_a,\beta_b)}
               e^{-\beta_a y_1-\beta_b y_2} \rd y_1\rd y_2,
\end{equation}
in which
\begin{eqnarray}
	 \Omega_{1,2}(y_1,y_2) &=& \frac{1}{\rd y_1\rd y_2}
   \int_{y_1<U_1(\omega)\le y_1+\rd y_1,\
y_2<U_2(\omega)\le y_2+\rd y_2}\rd\mathbb{P}(\omega)
\\
	&=& \frac{\partial^2}{\partial y_1\partial y_2}
\int_{U_1(\omega)\le y_1,\ U_2(\omega)\le y_2}\rd\mathbb{P}(\omega).
\end{eqnarray}
\end{subequations}
We note that the marginal distribution
\begin{equation}
    \int_{\mathbb{R}}  \frac{\Omega_{1,2}(y_1,y_2) }{Z_{1,2}(\beta_a,\beta_b)}
               e^{-\beta_a y_1-\beta_b y_2} \rd y_2
  = \frac{1}{Z_1(\beta_a)}\left(
     \int_{\mathbb{R}} \Omega_{1,2}(y_1,y_2)\rd y_2\right) e^{-\beta_a y_1}.
\end{equation}
This implies that
\begin{equation}
    \frac{1}{Z_1(\beta_a)} \int_{\mathbb{R}} \Omega_{1,2}(y_1,y_2)  \rd y_2
  =  \frac{1}{Z_{1,2}(\beta_a,\beta_b)}
     \int_{\mathbb{R}} \Omega_{1,2}(y_1,y_2)e^{-\beta_b y_2} \rd y_2.
\label{eq61}
\end{equation}
Since the rhs of (\ref{eq61}) is not a function of $\beta_b$, we
have the following equality:
\begin{equation}
  \frac{\partial\ln Z_{1,2}(\beta_a,\beta_b)}{\partial\beta_b}  =
            -\ \frac{\displaystyle \int_{\mathbb{R}} y_2 \Omega_{1,2}(y_1,y_2)e^{-\beta_b y_2} \rd y_2}{\displaystyle \int_{\mathbb{R}} \Omega_{1,2}(y_1,y_2)e^{-\beta_b y_2} \rd y_2}.
\end{equation}
Eq. \ref{eq61} can also be re-arranged into
\begin{subequations}
\begin{eqnarray}
 \frac{\displaystyle
     \int_{\mathbb{R}} \Omega_{1,2}(y_1,y_2)e^{-\beta_b y_2} \rd y_2 }{
        \displaystyle
     \int_{\mathbb{R}} \Omega_{1,2}(y_1,y_2) \rd y_2 } &=&
 \frac{Z_{1,2}(\beta_a,\beta_b)}{Z_1(\beta_a)},
\\
 \frac{\displaystyle
     \int_{\mathbb{R}} \Omega_{1,2}(y_1,y_2)e^{-\beta_by_2}
    \Big(e^{\beta_b y_2}\Big) \rd y_2 }{\displaystyle
     \int_{\mathbb{R}} \Omega_{1,2}(y_1,y_2)e^{-\beta_b y_2} \rd y_2 } &=&
 \frac{Z_1(\beta_a)}{Z_{1,2}(\beta_a,\beta_b)}.
\end{eqnarray}
\end{subequations}

{\bf\em Relative entropy between two random variables.}
The relative entropy between two measures
$\mu_1=\big(\mathbb{P}\oplus (-\beta_aU_1)\big)
\in\mathcal{M}$ and $\mu_2=\big(\mathbb{P}\oplus \big(-\beta_bU_2(\omega)\big)\big)\in\mathcal{M}$,
when transformed into the energy representation, is given by:
\begin{subequations}
\label{eq52}
\begin{eqnarray}
	\int_{\Omega}  \ln
       \left(\frac{\rd\mu_1}{\rd\mu_2}(\omega)\right)\rd\mu_1(\omega)  &=&
      \int_{\Omega} \frac{ e^{-\beta_a U_1(\omega)} }{Z_1(\beta_a)}
        \ln\left(\frac{Z_2(\beta_b)}{Z_1(\beta_a)}
            e^{-\beta_a U_1(\omega)+\beta_b U_2(\omega)}\right) \rd\mathbb{P}(\omega)
\nonumber\\
  &=&  \int_{\mathbb{R}}  \frac{ \Omega_{U_1}(h) e^{-\beta_a h} }{
           Z_1(\beta_a)}
          \ln\left( \frac{ \Omega_{U_2}(h)e^{-\beta_a h}  }{Z_1(\beta_a)  \Omega_{U_1}(h) }\right)\rd h
\\
   &+&   \beta_b\int_{\Omega}
  U_2(\omega) \left(\frac{ e^{-\beta_a U_1(\omega)} }{Z_1(\beta_a)}\right)
      \rd\mathbb{P}(\omega) \ + \ \ln Z_2(\beta_b).
\end{eqnarray}
\end{subequations}
Note that the first term in (\ref{eq52}b) again contains the
$\overline{U}_{2|U_1=h_1}$ that appeared in (\ref{e25}) and
(\ref{eq0037}).  It cannot be expressed in
terms of the energy represenations of $\mu_1$
and $\mu_2$.
Unless $g_2 = ag_1+b$, the two measures $\mu_1$ and
$\mu_2$, with densities $\rd\mu_1 = e^{g_1}\rd\mathbb{P}$
and $\rd\mu_2=e^{g_2}\rd\mathbb{P}$, do not share the same
restricted $\sigma$-algebra.

\subsection{Entropy divergence in the SoPMs}

Consider two probability measures $\mu_1,\mu_2\in\mathcal{M}$ in the SoPMs, with Radon-Nikodym derivatives w.r.t. $\mathbb{P}$ given by $f_1(\omega)$ and $f_2(\omega)$ respectively.  One can introduce the following divergence on $\mathcal{M}$:
\begin{equation}
d^2\big(\mu_1,\mu_2\big) = \int_{\Omega}\big(f_1(\omega) - f_2(\omega)\big)\left(\frac{\ln f_1(\omega) - \ln f_2(\omega)}{f_1(\omega) - f_2(\omega)}\right)\big(f_1(\omega) - f_2(\omega)\big)\mathbb{P}(\rd\omega).
\label{eq:metric}
\end{equation}

This divergence can also be rewritten as the sum of two non-negative terms in the form of relative entropy, a symmetrized version of the latter:
\begin{equation}
d^2\left(\mu_1,\mu_2\right) = \int_{\Omega} \ln\left(\frac{\rd\mu_1}{\rd\mu_2}(\omega)\right)\mu_1(\rd\omega) + \int_{\Omega}\ln\left(\frac{\rd\mu_2}{\rd\mu_1}(\omega)\right)\mu_2(\rd\omega).
\label{eq:metric2}
\end{equation}
From this second form, it is clear that $d$ is symmetric with respect to $\mu_1$ and $\mu_2$ and is zero if and only if $\mu_1 = \mu_2$ on $\mathcal{F}$.  This form also has the advantage of making it clear that $d$ is invariant with respect to the choice of an origin $\mathbb{P}$.  Note that, despite our notation, this quantity is not a metric because it does not satisfy the triangle inequality.  It is only a local metric.  That is, if $\mu_1$, $\mu_2$ and $\mu_3$ are sufficiently close together then $d(\mu_1, \mu_2) + d(\mu_2, \mu_3) \geq d(\mu_1, \mu_3)$.

{\bf\em Divergence in energy representation.}  If $\mu_1$ and $\mu_2$ are written in their respective energy representations, {\em i.e.}
$f_{E1}(y_1)=Z_1(\beta_a)\Omega_1(y_1)e^{-\beta_a y_1}$ and
$f_{E2}(y_2)=Z_2(\beta_b)\Omega_2(y_2)e^{-\beta_b y_2}$.
Then from Eq. \ref{eq:metric2}, we have
\begin{eqnarray}
    d^2\big(\mu_{1},\mu_{2}\big) &=& \beta_b\int_{\mathbb{R}}
              \left( \frac{\Omega_1(y_1)e^{-\beta_ay_1}}{Z_1(\beta_a)}
            \right)\overline{U}_{2|U_1=y_1}\ \rd y_1 -\beta_a\overline{U}_1(\beta_a) - \beta_b\overline{U}_2(\beta_b)
\nonumber\\
	&+&  \beta_a\int_{\mathbb{R}} \left(
               \frac{\Omega_2(y_2)e^{-\beta_by_2}}{Z_2(\beta_b)}\right)
    \overline{U}_{1|U_2=y_2}\ \rd y_2 .
\end{eqnarray}
There are three interesting special cases:

{\bf\em Different \mbox{\boldmath$\beta$}'s and same
\mbox{\boldmath$\Omega$}.}
If $\Omega_1(y)=\Omega_2(y)=\Omega(y)$,
\begin{equation}
     d^2\big(\mu_1,\mu_2\big) = \big(\beta_b-\beta_a\big) \Big(\overline{U}(\beta_a)-
           \overline{U}(\beta_b)\Big).
\label{d12-1}
\end{equation}

{\bf\em Different \mbox{\boldmath$\Omega$}'s and
same \mbox{\boldmath$\beta$}.}  With same $\beta_a=\beta_b=\beta$ but different $\Omega$'s,
\begin{subequations}
\label{d12-2}
\begin{eqnarray}
      d^2\big(\mu_1,\mu_2\big)  &=&
\beta\int_{\mathbb{R}}
              \left( \frac{\Omega_1(y_1)e^{-\beta y_1}}{Z_1(\beta)}
            \right) \Big[ \overline{U}_{2|U_1=y_1}-y_1\Big] \rd y_1
\\
	&+&  \beta\int_{\mathbb{R}} \left(
               \frac{\Omega_2(y_2)e^{-\beta y_2}}{Z_2(\beta)}\right)
   \Big[ \overline{U}_{1|U_2=y_2}-y_2\Big] \rd y_2
\\
	&=& \beta\Big( \overline{\mathcal{W}}_{12}(\beta)
           + \overline{\mathcal{W}}_{21}(\beta)\Big).
\end{eqnarray}
\end{subequations}
Here, following (\ref{0021}) and (\ref{def-work}),
we have identified the terms in (\ref{d12-2}a) and (\ref{d12-2}b)
as $\overline{\mathcal{W}}_{12}(\beta)$ and \
$\overline{\mathcal{W}}_{21}(\beta)$,
respectively.

{\bf\em Different \mbox{\boldmath$\Omega$}'s and
 \mbox{\boldmath$\beta$}'s.}
\begin{equation}
    d^2\big(\mu_1,\mu_2\big) =  \beta_b
          \overline{\mathcal{W}}_{12}(\beta_a)+
   \beta_a\overline{\mathcal{W}}_{21}(\beta_b) + \big( \beta_b -\beta_a\big) \Big( \overline{U}_1(\beta_a) - \overline{U}_2(\beta_b)
 \Big).
\label{eq0038}
\end{equation}
Eq. \ref{eq0038} implies an inequality that, being
different from (\ref{eq0031}) and (\ref{eq0034}), is based on
Massieu-Planck potential:
\begin{equation}
	 -\left(\frac{  \overline{\mathcal{W}}_{12} }{T_b}
     + \frac{ \overline{\mathcal{W}}_{21} }{T_a}  \right) \le
       \Big(\overline{U}_1-
           \overline{U}_2\Big)\left(\frac{1}{T_b} - \frac{1}{T_a}\right).
\label{eq0044}
\end{equation}

\subsection{Heat divergnece}

One can also introduce another related divergence on $\mathcal{M}$.  For fixed $\beta_a,\beta_b>0$, define:
\begin{eqnarray}
  d^2_{\beta}(\mu_1,\mu_2)  &=&  \frac{1}{\beta_a}
 \int_{\Omega} f_1(\omega)
    \ln\left(\frac{f_1(\omega)}{ f^{(\beta_a)}_2(\omega)}\right)
           \mathbb{P}(\rd\omega)
  + \frac{1}{\beta_b} \int_{\Omega}   f_2(\omega) \ln\left(\frac{f_2(\omega)}{ f^{(\beta_b)}_1(\omega)}\right)
    \mathbb{P}(\rd\omega)
\nonumber\\
	&=& \int_{\Omega}
      \left(\frac{e^{-\beta_aU_1(\omega)}}{Z_1(\beta_a)} -
           \frac{e^{-\beta_bU_2(\omega)}}{Z_2(\beta_b)}\right)
    \Big( U_2(\omega)-U_1(\omega)\Big) \mathbb{P}(\rd\omega)
\label{metric-2}\\
   &+&  \frac{1}{\beta_a}\ln\left(\frac{Z_2(\beta_a)}{Z_1(\beta_a)}\right)-
           \frac{1}{\beta_b}\ln \left(\frac{Z_2(\beta_b)}{Z_1(\beta_b)}\right),
\nonumber
\end{eqnarray}
in which
\begin{equation}
f_1(\omega) = \frac{e^{-\beta_a U_1(\omega)}}{Z_1(\beta_a)} \textrm{ and }
f_2(\omega) = \frac{e^{-\beta_b U_2(\omega)}}{Z_2(\beta_b)}
\end{equation}
are the densities of $\mu_1$ and $\mu_2$ with respect to $\mathbb{P}$ and
\begin{eqnarray}
     f^{(\beta_a)}_2(\omega) = \frac{e^{-\beta_aU_2(\omega)}}{
              Z_2(\beta_a)}, \
     f^{(\beta_b)}_1(\omega) = \frac{e^{-\beta_bU_1(\omega)}}{
              Z_1(\beta_b)}.
\end{eqnarray}
The same caveats as before apply: This is not a metric on $\mathcal{M}$ because it does not satisfy the triangle inequality, but it is a local metric in the sense that the triangle inequality is satisfied when all measures are sufficiently close together.  We shall call $d_{\beta}(\cdot,\cdot)$ in (\ref{metric-2}) the
{\em heat divergence}.  In terms of
\begin{equation}
              \mathcal{W}_{12}(\omega)=\frac{1}{\beta_a}
  \ln\left(\frac{e^{-\beta_a U_1(\omega)}}{e^{-\beta_a U_2(\omega)}}\right),\
     \mathcal{W}_{21}(\omega)=\frac{1}{\beta_b}
  \ln\left(\frac{e^{-\beta_b U_2(\omega)}}{e^{-\beta_b U_1(\omega)}}\right),
\end{equation}
we have
\begin{eqnarray}
          d^2_{\beta}(\mu_1,\mu_2) &=& \mathbb{E}^{\mu_1}
          \big[\mathcal{W}_{12}(\omega)\big]
+ \beta_a^{-1}
  \ln \mathbb{E}^{\mu_1}\Big[e^{-\beta_a\mathcal{W}_{12}(\omega)}\Big]    + \mathbb{E}^{\mu_2}
          \big[\mathcal{W}_{21}(\omega)\big]
\nonumber\\
	&+& \beta_b^{-1}
  \ln \mathbb{E}^{\mu_2}\Big[e^{-\beta_b\mathcal{W}_{21}(\omega)}\Big].
\label{e73}
\end{eqnarray}
Using the Jarzynski-Crooks relation from (\ref{eq0025}),  Eq. \ref{e73}
implies
\begin{equation}
   \overline{\mathcal{W}}_{12}(\beta_a)+
    \overline{\mathcal{W}}_{21}(\beta_b)
+  F_1(\beta_a) - F_2(\beta_a) + F_2(\beta_b)-F_1(\beta_b) \ge 0.
\end{equation}
This result generalizes Carnot's inequality.

\subsection{Infinitesimal entropy metric associated with
\mbox{\boldmath$\Delta\beta$}
}

Consider an infinitesimal change in
$\beta\to\beta+\Delta\beta$ and
corresponding $\rd\mu=e^{-\beta U}\rd\mathbb{P}\to \rd(\mu+\Delta \mu) = e^{-(\beta+\Delta\beta)U}\rd\mathbb{P}$.  Then we have
\begin{eqnarray}
    d^2\big(\mu,\mu+\Delta \mu\big) &=& (\Delta\beta)^2\int_0^{\infty}
             \frac{ \Omega(y) e^{-\beta y} }{Z(\beta)}
             \left[\left( \frac{\rd\ln Z}{\rd\beta}
         \right) +y \right]^2
              \rd y
\nonumber\\
&=& (\Delta\beta)^2\int_0^{\infty}
             \frac{ \Omega(y) e^{-\beta y} }{Z(\beta)}
             \Big(y - \mathbb{E}[U] \Big)^2
              \rd y
\nonumber\\
      &=& (\Delta\beta)^2\   \text{Var}\big[U\big].
\end{eqnarray}
This is a very important relation that connects
the {\em entropy divergnece} with {\em temerpature} and
{\em energy fluctuations}.   Furthermore, we have
\begin{equation}
  d^2\big(\mu,\mu+\Delta \mu\big)
  = (\Delta\beta)^2
         \left(-\frac{\rd^2\ln Z(\beta)}{\rd \beta^2}\right)
      \ =  \    (\Delta\beta)^2
         \left(\frac{\rd }{\rd \beta}\mathbb{E}\big[U\big] \right).
\end{equation}
The term inside $(\cdots)$ on the rhs is called the {\em heat capacity}
in thermodynamics.  Internal energy $\mathbb{E}[U]$ is
a ``slope'' and the $\text{Var}[X_{\beta}]$ is a curvature
of the ``potential function'' $-\ln Z(\beta)$.

\subsection{A mathematical remark}

{\bf\em Log-mean-exponential inequality and equality.}
We see that both entropy divergence in (\ref{eq:metric2})
and heat divergence in (\ref{e73}) are based on a very
general inequality involving the log-mean-exponential of
a random variable $\xi(\omega)$ \cite{qian-jpc-05}:
Jensen's inequality.
\begin{equation}
    \mathbb{E}\Big[\xi(\omega) \Big] +
    \beta^{-1}\ln \mathbb{E}\left[e^{-\beta\xi(\omega)}\right]
  \ge 0.
\label{equation80}
\end{equation}
In (\ref{eq:metric2}), the two $\xi$s are the information $\ln\frac{\rd\mu_1}{\rd\mu_2}(\omega)$ and
$\ln\frac{\rd\mu_2}{\rd\mu_1}(\omega)$;
and in (\ref{e73}), the two $\xi$s are the work
$\mathcal{W}_{12}(\omega)=\beta_a^{-1}\ln\frac{e^{-\beta_aU_1(\omega)}}{e^{-\beta_aU_2(\omega)}}$ and $\mathcal{W}_{21}(\omega)=\beta_b^{-1}\ln\frac{e^{-\beta_bU_2(\omega)}}{e^{-\beta_bU_1(\omega)}}$.  They are all different forms
of Radon-Nikodym derivatives.  In the entropy divergence, the
second, log-mean-exponential term in
(\ref{equation80}) is zero according to the Hatano-Sasa
equality.  In the heat divergence case, the same term
gives a Jarzynski-Crooks' free energy difference.

	Eq. \ref{equation80} should be recognized as
``mean internal energy minus free energy''.  Thus
it should be some kind of entropy:
\begin{equation}
  \mathbb{E}^{\mathbb{P}}\big[\xi(\omega) \big] +
    \beta^{-1}\ln \mathbb{E}^{\mathbb{P}}\Big[e^{-\beta\xi(\omega)}\Big]
   = \mathbb{E}^{\mathbb{P}}\left[\ln\left(
     \frac{\rd\mathbb{P}}{\ \rd\mathbb{P}'}(\omega) \right)\right],
\label{JensenE}
\end{equation}
in which $\mathbb{P}'=\mathbb{P}\oplus (-\beta\xi)$
is the affine sum of $\mathbb{P}$ and $(-\beta\xi)$.
Eq. \ref{JensenE} could be argued as the {\em fundamental
equation for isothermal processes} under a single temperature $T=\beta^{-1}$.
The implication of this interesting ``Jensen's equality'' to the
affine geometry of the SoPMs is currently being explored.

\section{Discussion}
\label{sec:5}

	It has been well established, through the work of
Gibbs, Carath\'{e}odory, and many others, that geometry
has a role in the theory of equilibrium thermodynamics
\cite{caratheodory,pogliani,salamon}. Classical
thermodynamics is not based on the theory of chance,
but there is no doubt that the notion of entropy has its root
in the theory of probability.  In the present work, we
propose that the space of probability measures as a
natural setting in which thermodynamic concepts can be
established logically.  In particular, an affine structure is
natually related to the canonical probability distribution
studied by Boltzmann and Gibbs in their statistical
theories, and almost all thermodynamic potentials are
different forms of Radon-Nikodym derivatives associated
with {\em changes of measures}.  Even
the fundamental equation of nonequilibrium
therodynamics, together with the distinctly nonequilibrium
notion of entropy production, naturally emerge.

	Statistical mechanics, as a scientific theory,
differs from Kolmogorov's axiomatic theory of probability in
one essential point: The latter demands
a complete probability space and a normalized probability
measure, while in the former every probability distribution is
a {\em conditioned probability} under many known and
unknown conditions.  More importantly, the
probability of the conditions, themselves as random events,
are usually not knowable.  In
the theory of the space of measures, we see that one
mechanical system with a given energy function
$U(\omega)$ corresponds to a straight line, and the fixing
of the origin in $\mathcal{M}$ in terms of $\mathbb{P}$
or the normalization in terms of $Z(\beta)$ [which translates
to the arbitary constant in $U(\omega)$]  amount to the
idea of gauge fixing.  Thermodynamic work then arises
in the rotation from $U_1(\omega)$ to
$U_2(\omega)$.  In the theory of probability, associated
with any ``change'' is a {\em change of measure}:
Radon-Nikodym derivatives simply provide the calculus
to quantity the {\em fluxion}!  In Newtonian mechanics,
change in space is absolute; but in probability, it is a
complex matter, and it is all relative.

	The probability theory of large deviations is now a
recognized mathematical foundation for statistical
thermodynamics \cite{ellis,touchette,ge-qian-16}.  Such a
theory is concerned with the deterministic
{\em thermodynamic limit}.  In
Sec. \ref{sec:3.6}, we see that the combination of
our theory and a deterministic limit gives rise to the
concept of {\em macroscopic equations of state} in classic
thermodynamics \cite{pauli}.

Equilibrium mean internal energy $\overline{U}(\beta)$ depends on both the intrinsic properties of a system and its external environment.
This is most clearly shown through the canonical distribution
that is determined by $U(\omega)$ and $\beta$.
The decomposition in Eq. \ref{eq13}, a simple example of
the much more general (\ref{JensenE}), connects the internal energy
with ``work'' and ``heat'', or the ``usable energy'' and
``useless energy'', or entropy production and entropy change.
These are all just different interpretations under different perspectives.

\section*{Acknowledgements}
We thank Yu-Chen Cheng
and Ying-Jen Yang for many helpful discussions,
and Professors Jin Feng (University of Kansas) and Hao Ge (Peking
University) for advices. L.H. acknowledges the financial supports from the National Natural Science Foundation of China (Grants 21877070) and Tsinghua University Initiative Scientific Research Program (Grants 20151080424). H.Q. acknowledges the Olga Jung Wan Endowed Professorship for support.

\end{document}